# Interface states in polariton topological insulators

Yiqi Zhang[1,*], Yaroslav V. Kartashov[2], and Albert Ferrando[3]

[1]*Key Laboratory for Physical Electronics and Devices of the Ministry of Education and Shaanxi Key Lab of Information Photonic Technique, The School of Electronic and Information Engineering, Xi'an Jiaotong University, Xi'an 710049, China*
[2]*Institute of Spectroscopy, Russian Academy of Sciences, Troitsk, Moscow Region 108840, Russia*
[3]*Departament d'Òptica, Interdisciplinary Modeling Group, InterTech, Universitat de València, 46100 Burjassot (València), Spain*
[*]*Corresponding author: zhangyiqi@mail.xjtu.edu.cn*

**Abstract:** We address linear and nonlinear topological interface states in polariton condensates excited at the interface of the honeycomb and Lieb arrays of microcavity pillars in the presence of spin-orbit coupling and Zeeman splitting in the external magnetic field. Such interface states appear only in total energy gaps of the composite structure when parameters of the honeycomb and Lieb arrays are selected such that some topological gaps in the spectrum of one of the arrays overlap with topological or nontopological gaps in the spectrum of other array. This is in contrast to conventional edge states at the interface of periodic topological and uniform trivial insulators, whose behaviour is determined exclusively by the spectrum of topological medium. The number of emerging interface states is determined by the difference of Chern numbers of overlapping gaps. The illustrative examples with one or two coexisting unidirectional interface states are provided. The representative feature of the system is the possibility of wide tuning of concentration of power of the interface states between two limiting cases when practically all power is concentrated either in Lieb or in honeycomb array. Localization of the interface states and their penetration depth into arrays drastically vary with Bloch momentum or upon modification of the amplitude of the interface state in the nonlinear regime. We illustrate topological protection of the interface states manifested in the absence of backscattering on interface defects, and study their modulation instability in the nonlinear regime. The latter leads to formation of quasi-solitons whose penetration into different arrays also depends on Bloch momentum. In addition, we discuss the impact of losses and coherent pump leading to bistability of the interface states.

**Keywords:** Microcavities; Polaritons; Topological insulators; Topological interface state; Quasi-solitons

## 1. Introduction

Localization and propagation of waves of various physical nature at the interface of materials with different physical properties is a topic of continuously renewed interest. The problem of formation of such waves appears in diverse areas of science, including acoustics, hydrodynamics, radio-physics, physics of matter waves, and optics. Waves arising and propagating along the interface of two materials are called surface waves. Their properties at the interfaces of uniform materials are well-known, but when such waves form at the interfaces of spatially inhomogeneous, in particular, periodic materials with bulk spectra exhibiting allowed and forbidden energy bands, they acquire completely new features that are under active investigation nowadays. The problem of formation of interface states naturally appears in photonic crystals, see reviews on resonant linear interface states in such crystals [1,2] and in shallow optical lattices/waveguide arrays [3,4], where nonlinearity can be used to localize light at the interface of uniform medium and one- [5] or two-dimensional [6,7] array or at the interface between two dissimilar arrays [8-12]. Even though surface waves in topologically trivial materials reflect rich structure of the bandgap spectra of periodic media placed in contact, they are not protected by the system topology and, therefore, are sensitive to perturbations in periodic structure.

Completely different physical scenario is realised when waves at the interface of two media appear because they feature different topology. Such waves are said to be protected by the topology and due to this they show remarkable resistance to local deformations of the underlying materials. Usually topological edge states appear when one of the materials forming the interface is periodic and possesses specific degeneracies in its spectrum. Edge states at the interfaces of such materials (including those with honeycomb and hexagonal structure) may be created by introducing deformations into underlying lattices [13], by varying spacing between lattice sites across the interface, introducing detuning between sublattices, changing orientation of anisotropic elements placed in the lattice nodes, and realization of anisotropic coupling between different lattice sites. They have been suggested and observed in electromagnetic systems, including photonic crystals [14-16], periodic metamaterial structures [17,18], shallow waveguide arrays [19], for acoustic [20] and elastic [21] waves, polaritons [22], as well as for electronic states in two-dimensional materials [23]. All states mentioned above belong to special class that does not require breakup of time-reversal symmetry in governing evolution equations for its existence.

Breakup of time-reversal symmetry in the system with specific degeneracies in the spectrum leads to opening of topological gap with unique unidirectional topological states in it [24,25]. The direction of propagation of such states can be reversed by swapping materials forming the interface or, in systems where time-reversal symmetry is broken by the external magnetic field, by changing the direction of this field. Such states represent very robust topologically protected and, most importantly, travelling excitations that generally cannot be destroyed by perturbations with energies smaller than the width of the topological gap, where they form. They, or their analogues, have been proposed and observed in different physical systems, including acoustic [26], photonic [27-37], and optoelectronic [38-42] ones. Including nonlinear effects substantially enriches the behaviour of the unidirectional topological states, leading to, e.g., nonlinearity-mediated inversion of the propagation direction of the edge states [43,44], development of modulation instability [45,46], formation of topological edge solitons [47-50], and bistability effects [51].

Commonly considered configuration in systems with broken time-reversal symmetry assumes the interface between periodic and uniform material. Thus, rich physics emerging when interface is instead created by two periodic media with different symmetry and bandgap spectra and unidirectional states at such interface remains unexplored, to the best of our knowledge. Here we employ polaritonic system, where edge states were recently observed experimentally [41], to demonstrate new possibilities for control of topological edge states at the interface of honeycomb and Lieb arrays of microcavity pillars located in the external magnetic field. Such arrays are available for experimental exploration [52-56]. We consider the interface of two arrays with different symmetries and number of elements per unit cell that cannot be transformed into each other using continuous deformations. Broken time-reversal symmetry by virtue of simultaneously present spin-orbit coupling and Zeeman splitting in the external magnetic field, leads to the appearance of the in-gap topological interface states, when two topological or topological and nontopological energy gaps in honeycomb and Lieb arrays overlap. The number of such states is determined by the difference of Chern numbers of overlapping gaps. We illustrate topological protection for such interface states, their unusual localization properties, and introduce interface quasi-solitons.

## 2. The model

The evolution of the spinor wavefunction $\Psi = (\psi_+, \psi_-)^T$ describing polariton condensate in the potential landscape, created by two arrays of the microcavity pillars with honeycomb and Lieb symmetry placed in contact, is governed by the system of dimensionless coupled Schrödinger equations [38,48]:

$$i\frac{\partial \psi_\pm}{\partial t} = -\frac{1}{2}\left(\frac{\partial^2}{\partial x^2} + \frac{\partial^2}{\partial y^2}\right)\psi_\pm + \beta\left(\frac{\partial}{\partial x} \mp i\frac{\partial}{\partial y}\right)^2 \psi_\mp \pm \Omega\psi_\pm + \mathcal{R}(x,y)\psi_\pm + (|\psi_\pm|^2 + \sigma|\psi_\mp|^2)\psi_\pm. \quad (1)$$

Here $\psi_\pm$ are the complex wavefunctions of the spin-positive and spin-negative polaritons in circular polarization basis; the parameter $\beta$ accounts for the spin-orbit coupling effect stemming from TE-TM energy splitting in the microcavity [38,57]; the terms $\sim \Omega$ account for the Zeeman energy splitting in the external magnetic field. Nonlinear terms account for the repulsion between polaritons with the same spin, while $\sigma = -0.05$ corresponds to the weak cross-spin attraction [58]. Our composite array of microcavity pillars is the union of honeycomb and Lieb arrays with straight interface between them that is parallel to the $y$-axis. Corresponding potential landscape created by the microcavity pillars $\mathcal{R}(x,y) = \mathcal{R}_h(x,y) + \mathcal{R}_l(x,y)$ is described by the functions:

$$\mathcal{R}_h(x,y) = -\sum_{n,m,x \leq x_{in}} p_h \mathcal{Q}(x-x_n, y-y_m),$$
$$\mathcal{R}_l(x,y) = -\sum_{p,q,x > x_{in}} p_{a,b} \mathcal{Q}(x-x_p, y-y_q),$$

where individual Gaussian potential wells $\mathcal{Q} = \exp[-(x^2+y^2)/d^2]$ are placed in the nodes of the honeycomb grid $x_n, y_m$ at $x < x_{in}$, and in the nodes of Lieb grid $x_p, y_q$ at $x > x_{in}$, with $x_{in}$ being the interface position [see Fig. 1(d) for $-\mathcal{R}$ shape and Fig. 1(e) for schematic representation]. The separation between neighbouring potential wells is given by $a$ in the honeycomb array and by $3^{1/2}a/2$ in the Lieb array. To allow control of relative positions of energy gaps in two arrays we set $p_h$ as a potential depth in the honeycomb array [Fig. 1(b)], and assume that Lieb array contains wells with different depths $p_a$ and $p_b$, $p_b > p_a$ [Fig. 1(c)]. This composite structure is periodic in the $y$ direction: $\mathcal{R}(x,y) = \mathcal{R}(x,y+Y)$ with period $Y = 3^{1/2}a$.

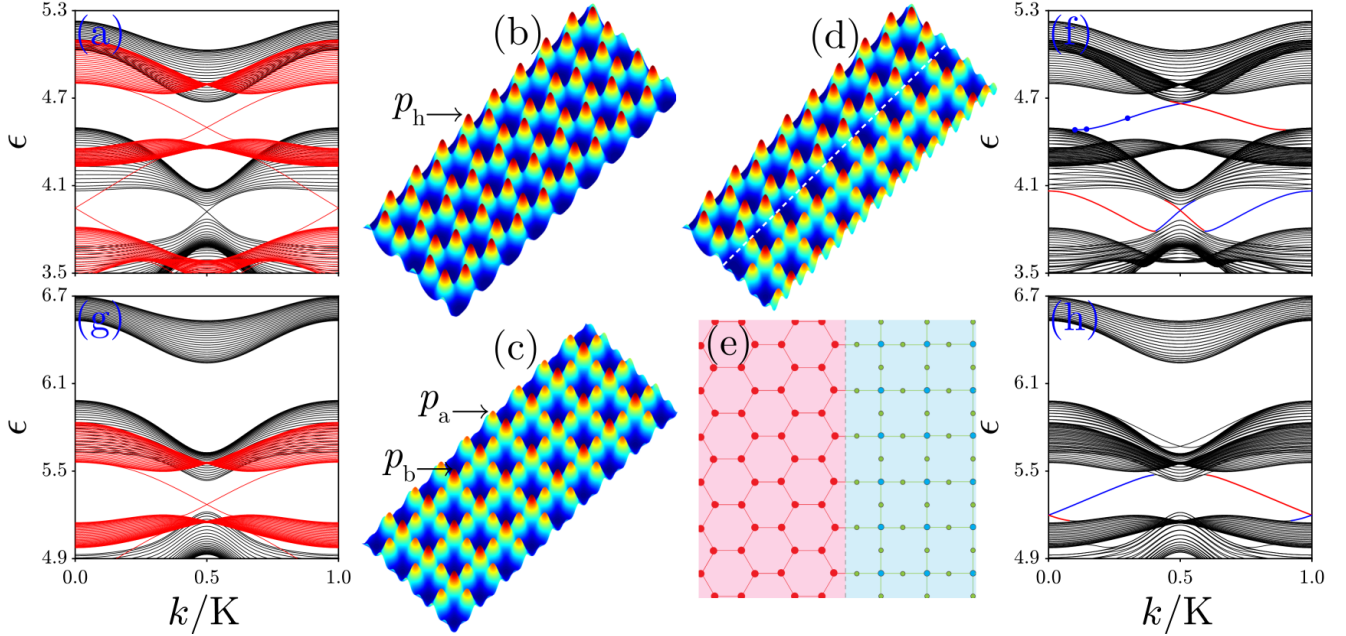

Fig. 1. (a) Band structure of the truncated in $x$ honeycomb array with $p_h = 10$ (red curves) and that of the truncated Lieb array with $p_a = 8$ and $p_b = 9.7$ (black curves). Band structures are calculated for zigzag-zigzag boundaries for honeycomb array and bearded-bearded boundaries for Lieb array. For these parameters first finite gaps in eigenvalue spectrum of two arrays overlap with each other. (b) $-\mathcal{R}_\text{h}$ and (c) $-\mathcal{R}_\text{l}$ profiles with indication of depths of the potential wells. (d) $-\mathcal{R}$ profile for composite array with dashed line indicating interface and (e) corresponding schematic representation, indicating deeper (blue) and shallower (green) potential wells in the Lieb array. (f) Band structure of the composite honeycomb-Lieb array for the same $p_\text{h}, p_\text{a,b}$ values as in (a). To make band structure symmetric, the honeycomb array slab was placed between two Lieb arrays. Blue and red curves correspond to the interface states on the right and left edges of the slab, respectively. (g),(h) The same as in (a),(f), but for $p_\text{h} = 11.25$, $p_\text{a} = 10$, and $p_\text{b} = 12.1$, when first gap of the honeycomb array overlaps with the second gap of the Lieb array. All quantities are plotted in dimensionless units.

Spatial coordinates in (1) are normalized to the characteristic distance $L$, all energy parameters (such as potential depth and Zeeman splitting) are normalized to the characteristic energy $\epsilon_0 = \hbar^2/mL^2$, where $m$ is the effective polariton mass, while evolution time is normalized to $\hbar\epsilon_0^{-1}$, see [48] for details. We omit losses in our model that are typical for polariton condensates [59,60], because the existence of interface states is not connected with them and because proper pumping can be used to dramatically increase the lifetime of the interface states.

It should be stressed here that the main reason why we consider here the interface of two different lattices – honeycomb and Lieb – is the possibility to obtain in this structure more than one edge state per interface. If the symmetry of two lattices were the same, the interface between two lattices could only be created (without changing the direction of the magnetic field) by changing their depths or by deforming one of them, but in those structures the arrangement of Chern numbers for gaps may be identical, so it may be not easy to find configuration with more than one edge state per interface. On the other hand, at the usual interface between uniform medium and the lattice, the edge state always expands more into lattice region, but not into uniform medium, thus limiting the possibilities for control of localization of the edge states.

## 3. Linear topological interface states

As it was shown in [38,48] the necessary ingredients for appearance of the unidirectional topological edge states in truncated arrays of microcavity pillars include simultaneous action of spin-orbit coupling and Zeeman splitting in the magnetic field, together with the presence of specific degeneracies in the spectrum of the array, for example, in the form of Dirac points. Our arrays are selected in such a way that such degeneracies can present in different gaps in spectra of both arrays depending on their depths $p_\text{h}$ and $p_\text{a,b}$ (further we set $p_\text{b} > p_\text{a}$ to be able to match the spectral gaps in two arrays, since this is required for localization of interface states at both sides of the interface, as explained below). Simultaneous action of spin-orbit coupling effect and Zeeman splitting that breaks time-reversal symmetry in Eq. (1) can open topological gaps that can coexist with nontopological gaps, appearing for example in Lieb array due to introduced difference in potential depths $p_\text{a,b}$. To illustrate this we first consider linear spectra of separate honeycomb and Lieb arrays that will be used afterwards for construction of the composite structure. We assume that such arrays are truncated along the $x$-axis, so that their linear modes are Bloch waves $\psi_\pm(x,y,t) = u_\pm(x,y)\exp(iky + i\epsilon t)$ periodic in $y$ and localized in $x$: $u_\pm(x,y) = u_\pm(x,y+\text{Y})$ and $u_\pm(x\to\pm\infty, y) = 0$, where $k$ is the Bloch momentum and $\epsilon$ is the energy. The latter is a periodic function of $k$ with a period $\text{K} = 2\pi/\text{Y}$. Resulting linear eigenvalue problem

$$\epsilon u_\pm = (1/2)[\partial^2/\partial x^2 + (\partial/\partial y + ik)^2]u_\pm - \mathcal{R}_\text{h,l}(x,y)u_\pm \\ -\beta[\partial/\partial x \mp i(\partial/\partial y + ik)]^2 u_\mp \mp \Omega u_\pm, \quad (2)$$

obtained after substitution of the wavefunction $(\psi_+, \psi_-)^\text{T}$ into Eq. (1), with truncated honeycomb $\mathcal{R}_\text{h}$ or Lieb $\mathcal{R}_\text{l}$ potential was solved numerically using plane-wave expansion method. In simulations we use representative parameter values $\beta = 0.3$, $\Omega = 0.5$, $a = 1.4$, $d = 0.5$ allowing us to obtain sufficiently wide topological gaps in the spectrum.

Black and red lines in Fig. 1(a) show, respectively, linear spectra in the form of energy-momentum dependencies $\epsilon(k)$ for purely Lieb or purely honeycomb arrays, when they are truncated along $x$ (here we consider honeycomb array with zigzag-zigzag edges and Lieb array with bearded-bearded edges). This type of truncation is selected in the view of the similar truncation that will be used at the interface of composite array, see Figs. 2(a) and 1(e). The depths of the potential wells in Fig. 1(a) are adjusted such as to achieve overlap of the first and second finite gaps in the spectrum of honeycomb and Lieb arrays. One can clearly see the presence of the topological edge states (one per interface) in the spectrum of honeycomb array in both first and second gaps (such states bifurcate from top band around $k = \text{K}/3$ and $k = 2\text{K}/3$ momentum values). In contrast, no edge states are observed in the first nontopological gap of truncated Lieb array, but they are visible around $k = \text{K}/2$ in the second topological gap of this array. This means that the first gap in this array is actually opened due to introduced difference $p_\text{b} - p_\text{a}$ between depths of the potential wells and it persists even at $\beta, \Omega = 0$, while the second gap opens due to breakup of time-reversal symmetry.

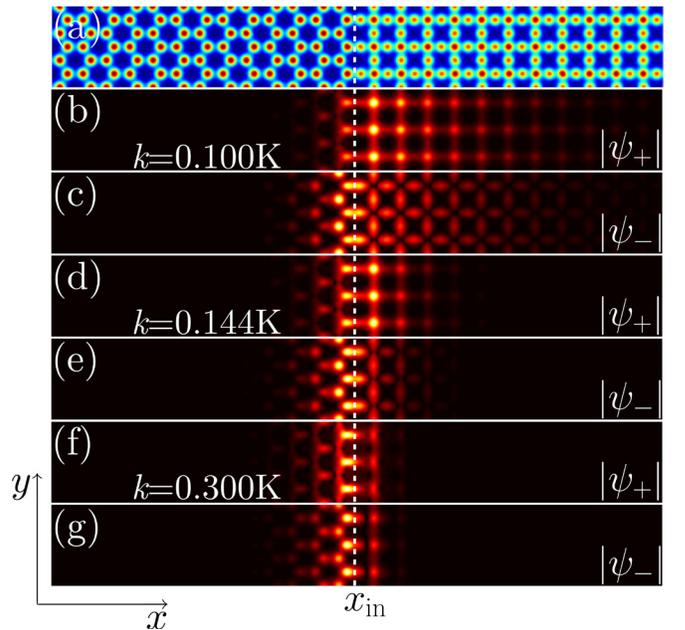

Fig. 2. (a) The array configuration and (b)-(g) interface states at different Bloch momentum values indicated in the panels. Both spin-positive and spin-negative wavefunction components are shown within the window $0 \le x \le 55$ and $-3\text{Y}/2 \le y \le +3\text{Y}/2$. White dashed line

indicates interface position. All quantities are plotted in dimensionless units.

Remarkably, when honeycomb and Lieb arrays with above parameters are used to form composite array shown in Fig. 1(d), its spectrum (calculated using Eq. (2), but now with composite potential $\mathcal{R}$) clearly shows the presence of the interface states localized in both arrays in *total* gaps of this structure [Fig. 1(f)]. Notice that to make this spectrum symmetric we placed wide honeycomb slab between two Lieb arrays, rather than considering single interface, so interface states in Fig. 1(f) appear in pairs with opposite slopes $\partial\epsilon/\partial k$. In this figure blue curves corresponds to states on the right interface, red curves correspond to states on the left interface, and black curves show bulk modes. Here we use the term *interface* states to stress that they form between two dissimilar periodic media, see Fig. 2 for examples of such states from the first total gap. The domain of existence of the interface states in momentum $k$ may notably differ from those in constituting arrays: by comparing Figs. 1(f) and 1(a) one can see that interface states in the top gap now connect band regions with dispersion inherited from Lieb [black curves in Fig. 1(a)], rather than from honeycomb [red curves in Fig. 1(a)] array. This suggests that close to gap edges such edge states may penetrate much stronger into Lieb array than to the honeycomb one (see discussion below). Moreover, slight variation of parameters may lead to situation when interface states will connect dispersion regions inherited from *different* arrays, in which case they will strongly extend into *different* arrays close to the upper or lower edges of total topological gap.

Remarkably, Fig. 1(f) reveals the presence of different number of interface states in the first and second gaps of composite structure: there is *one* state per interface in the first gap and *two* states per interface in the second gap. The existence of the interface states in composite array can be anticipated by considering *gap* Chern numbers $\mathcal{C}_{h,l}^\alpha$ for corresponding bulk honeycomb and Lieb arrays, defined as a sum of Chern numbers for all bands lying above selected gap [32,33]. The superscript $\alpha$ in $\mathcal{C}_{h,l}^\alpha$ is an integer denoting the index of the spectral gap. Standard procedure of calculation of the topological invariants [50] yields for the parameters of Fig. 1(a) gap Chern numbers $\mathcal{C}_h^{\alpha=1} = -1$ and $\mathcal{C}_h^{\alpha=2} = -1$ in the honeycomb array, and $\mathcal{C}_l^{\alpha=1} = 0$ and $\mathcal{C}_l^{\alpha=2} = +1$ in the Lieb array. When two arrays are placed in contact, the number of the topological states per interface existing in selected gap of the composite structure is determined by the modulus of difference of the gap Chern numbers $\left|\mathcal{C}_h^\alpha - \mathcal{C}_l^\gamma\right|$ [24], where $\alpha, \gamma$ are the indices of the overlapping gaps. This expression predicts *one* interface state in the first finite gap and *two* states in the second gap of composite structure visible in Fig. 1(f) (there can be more edge states in still lower gaps, but we do not consider them here for simplicity). Thus, in polariton condensate one can realize rare configuration with more than one topological state per interface by composing two periodic structures with different symmetries and properly selected depths of the potential wells. Moreover, even if there are no topological states in the gap of one of the arrays, they appear in approximately the same energy interval in composite structure, if other array admits them in suitable gap. Notice, that if the topological gap of the honeycomb array overlaps with band in Lieb array, one gets states that are confined at honeycomb side of the interface but are delocalized at the Lieb side. The depths of potential wells $p_h$ and $p_{a,b}$ can be tuned such as to achieve overlap of the first gap in honeycomb array with the second gap of the Lieb array [Fig. 1(g)]. However, for parameters of Fig. 1(g) one has $\mathcal{C}_h^{\alpha=1} = -1$ and $\mathcal{C}_h^{\alpha=2} = -1$, but both gaps in the Lieb array are nontopological now: $\mathcal{C}_l^{\alpha=1} = 0$, $\mathcal{C}_l^{\alpha=2} = 0$. In this case only one state per interface appears in the second gap of the composite structure

[Fig. 1(h)]. Such interface states have different structure in the Lieb array and they exist in different interval of Bloch momenta in comparison with interface states from Fig. 1(f). The part of the interface state branch located close to the top of the topological gap corresponds to states that are concentrated more in the Lieb array, the part of the branch close to the bottom of the gap corresponds to states extending more into honeycomb array.

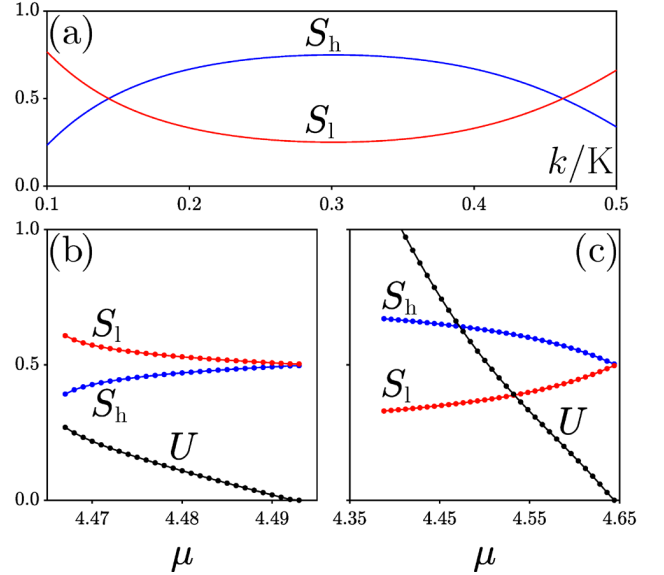

Fig. 3. (a) Norm sharing between honeycomb $S_h$ and Lieb $S_l$ arrays versus Bloch momentum for linear interface states corresponding to the top blue curve in Fig. 1(f). (b,c) Norm sharing versus energy $\mu$ for the nonlinear interface states at (b) $k = 0.144\mathrm{K}$ [left crossing point in panel (a)] and (c) $k = 0.462\mathrm{K}$ [right crossing point in panel (a)]. All quantities are plotted in dimensionless units.

One of the important advantages of our system with interface between two periodic structures in comparison with usual setting, where topological material contacts with a uniform one, is the possibility to tune penetration depth of the interface states into different arrays. In our case this penetration depth and norm (or power) distribution between arrays strongly depend on the Bloch momentum $k$. Examples of the interface states corresponding to the blue circles in Fig. 1(f) are shown in Figs. 2(b)-(g). To determine the fraction of norm of these interface states contained in different arrays we introduce the quantities:

$$U_h = \int_{\mathcal{R}_h} dx \int_{-Y/2}^{+Y/2} \boldsymbol{\Psi}^\dagger \boldsymbol{\Psi}\, dy,$$
$$U_l = \int_{\mathcal{R}_l} dx \int_{-Y/2}^{+Y/2} \boldsymbol{\Psi}^\dagger \boldsymbol{\Psi}\, dy, \qquad (3)$$

where $\boldsymbol{\Psi} = (\psi_+, \psi_-)^\mathrm{T}$ and integration in $y$ is performed over one $y$-period of the wave, while integration in $x$ is performed over regions occupied by the honeycomb $\mathcal{R}_h$ or Lieb $\mathcal{R}_l$ arrays; $U = U_h + U_l$ is the total norm per period. It is convenient to introduce norm sharing as $S_{h,l} = U_{h,l}/U$. Interface state from Figs. 2(b),(c) is mostly concentrated within Lieb array, hence $S_l \gg S_h$. This is consistent with the fact that this interface state is taken close to the edge of the topological gap, where interface state curve connects with the part of the band inherited from Lieb array. By adjusting Bloch momentum one can achieve the situation shown in Figs. 2(d),(e), where norms and penetration depths in both arrays are almost equal, $S_l \approx S_h$. Finally, at $k = 0.3\,\text{K}$, close to the center of the gap, the interface state penetrates deeper and has most of its norm concentrated in the honeycomb array, $S_l \ll S_h$ [Figs. 2(f),(g)]. Notice that since this interface state connects on the top of the gap with the band that inherited dispersion from Lieb array, further increase of $k$ again leads to growth of the fraction of norm $S_l$ contained in the Lieb array. Variation in norm sharing with Bloch momentum for linear interface state from the top gap in Fig. 1(f) is shown in Fig. 3(a). Notice that there are two points, $k = 0.144\text{K}$ and $k = 0.462\text{K}$, where norm is distributed equally between two arrays, i.e. $S_h \approx S_l$.

To confirm that interface states obtained here are indeed topologically protected we remove two pillars from the array that are located in the vicinity of interface [Fig. 4(a)] and study interaction of the topological state from the top blue branch in Fig. 1(f) with such a defect. Sufficiently wide Gaussian envelope was superimposed on the input wave, as shown in Fig. 4(b). This state moves in the negative direction of the $y$-axis. One can see that it passes the defect [Figs. 4(c),(d)] without observable backscattering and radiation into honeycomb or Lieb regions that is a clear signature of topological protection. Similar protection was observed for interface states from Fig. 1(h), residing in the second gap of the composite array.

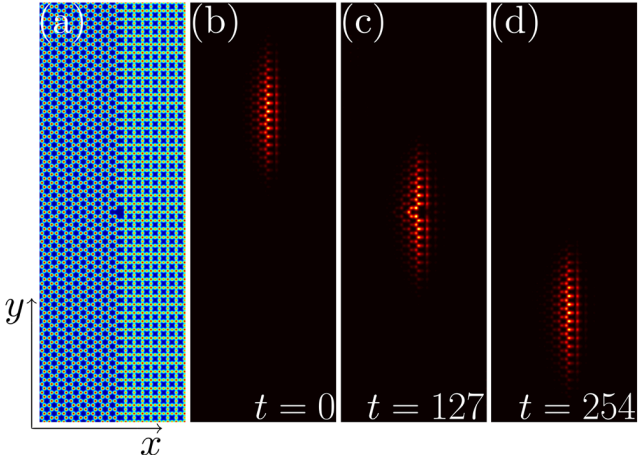

Fig. 4. (a) $-\mathcal{R}$ profile with a defect on the interface. (b)-(d) Interaction of the interface state with $k = 0.3\text{K}$ and broad Gaussian envelope with a defect. Only $|\psi_-|$ distribution is shown in different moments of time within $1.5 \leq x \leq 43.5$ and $-25\,\text{Y} \leq y \leq +25\,\text{Y}$ windows. All quantities are plotted in dimensionless units.

## 4. Nonlinear topological interface states

Nonlinearity also strongly affects localization of the interface states and changes their structure. Nonlinear interface states can be found from Eq. (1) in the form $\psi_\pm(x,y,t) = u_\pm(x,y)\exp(iky + i\mu t)$, where $\mu$ is the energy, while $y$-periodic functions $u_\pm(x,y) = u_\pm(x, y+\text{Y})$ satisfy the equation similar to (2), but with nonlinear terms included:

$$\mu u_\pm = (1/2)[\partial^2/\partial x^2 + (\partial/\partial y + ik)^2]u_\pm - \mathcal{R}(x,y)u_\pm \\ -\beta[\partial/\partial x \mp i(\partial/\partial y + ik)]^2 u_\mp \mp \Omega u_\mp - (|u_\pm|^2 + \sigma|u_\mp|^2)u_\pm. \quad (4)$$

This equation was solved using Newton's iterative method. Such nonlinear modes bifurcate from linear interface states. In the point of bifurcation the energy $\mu$ coincides with the eigenvalue $\epsilon$ of the linear interface state, while amplitude of the nonlinear state vanishes. Norm $U$ per $y$-period and amplitude of the nonlinear interface states increase with decrease of $\mu$. Usually the $x$-width of the state also increases with decrease of $\mu$. We calculated $U(\mu)$ dependencies for parameters used in Fig. 1(f) for two representative momentum values $k = 0.144\text{K}$ [Fig. 3(b)] and $k = 0.462\text{K}$ [Fig. 3(c)] for which norm sharing between two arrays is equal in the linear limit. Nonlinearity substantially affects norm sharing, see blue and red curves with circles showing fraction of the norm concentrated in honeycomb $S_h$ and Lieb $S_l$ arrays, respectively. While for smaller momentum value the nonlinear interface state extends more into Lieb array with increase of its norm, for larger momentum the tendency is reversed and the interface state extends more into honeycomb array. In both cases one observes delocalization (typically at one side of the interface) when $\mu$ reaches the border of the topological gap. For certain Bloch momentum values $k$ selected in between values mentioned above one can observe transition between $S_l > S_h$ and $S_l < S_h$ cases upon variation of energy $\mu$. Representative profiles of the nonlinear interface states illustrating such a transition at $k = 0.18\text{K}$ are shown in Fig. 5. While state in Figs. 5(a),(b) has $S_l > S_h$ (i.e. it is concentrated more in the Lieb array), its counterpart with smaller norm from Figs. 5(c),(d) corresponds to $S_l < S_h$ (i.e. it is concentrated more in honeycomb array). Thus, norm sharing in interface states can be effectively controlled also by nonlinear effects.

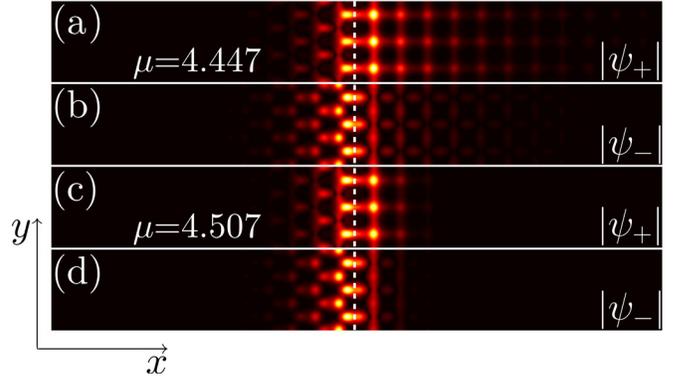

Fig. 5. Nonlinear interface states from the same family with $S_l > S_h$ (a),(b), and $S_l < S_h$ (c),(d). Bloch momentum $k = 0.18\text{K}$, while corresponding $\mu$ values are shown in the panels. The distributions are shown within $0 \leq x \leq 55$ and $-3\text{Y}/2 \leq y \leq +3\text{Y}/2$ windows. All quantities are plotted in dimensionless units.

## 5. Modulational instability and interface quasi-solitons

Even though dominating nonlinearity in polariton insulator is repulsive, modulational instabilities are not excluded and they may occur for nonlinear interface states, provided that second-order dispersion for corresponding branch of interface states has proper sign. The second-order dispersion coefficient $\partial^2\epsilon/\partial k^2$ for two interface states from the top gap in Fig. 1(f) is show in Fig. 6(a) as a function of momentum $k$ [we use the same colour scheme as in Fig. 1(f) to denote different branches]. One can see that there exist broad momentum intervals, where dispersion coefficient $\partial^2\epsilon/\partial k^2 > 0$ (effective polariton mass is negative) and the development of modulational instability is possible. Indeed, propagation of slightly perturbed nonlinear inter-face states with corresponding values of momentum $k$ reveals their fragmentation into sets of bright spots along the interface. Representative evolution of peak amplitudes $a_\pm = \max|\psi_\pm|$ of spin-positive and spin-negative wavefunction components in perturbed nonlinear interface state with $k=0.25$ and $\mu=4.53$ [corresponding dispersion coefficient is indicated by the blue dot in Fig. 6(a)] is shown in Fig. 6(b). Spatial distributions of the wavefunction modulus in strongest $\psi_-$ component corresponding to green dots in Fig. 6(b) are depicted in Fig. 7(a). One can see that at sufficiently large times nonlinear edge state is fragmented into sets of weakly radiating localized bright spots. These spots can be considered as precursors for the formation of the interface quasi-solitons. We term such states quasi-solitons because they are metastable topologically protected objects travelling along the lattice interface, but experiencing very slow reshaping due to radiation into the bulk and higher-order dispersion, in contrast to usual solitons that stay invariable in the course of propagation.

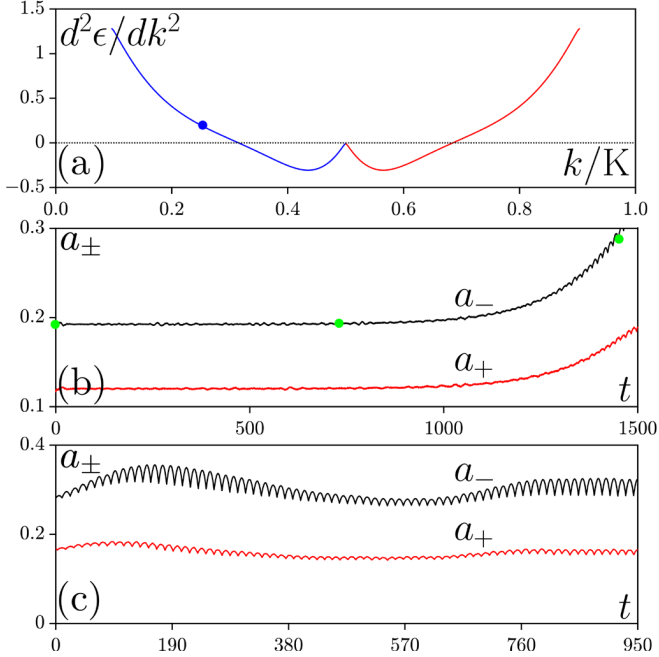

Fig. 6. (a) Second-order dispersion coefficient $\partial^2\epsilon/\partial k^2$ for interface states from the top gap of Fig. 1(f) versus momentum $k$. The colour scheme is the same as in Fig. 1(f). (b) Peak amplitudes of the two components $\psi_\pm$ in perturbed nonlinear interface state with $k=0.25K$ and $\mu=4.53$ versus time. (c) Peak amplitudes of components of quasi-soliton, constructed using bright spot marked with green circle in Fig. 7(a), versus time. All quantities are plotted in dimensionless units.

Indeed, if one isolates one of these spots, for example the spot indicated by the green circle in the pattern at $t=1450$, by imposing localized envelope on corresponding wave profile and uses such a wavepacket as an input in the evolution Eq. (1), one observes [Fig. 7(b)] immediate formation of stable quasi-soliton moving along the interface of two arrays with small amplitude oscillations [Fig. 6(c)] and practically without radiation into the bulk of arrays. The velocity of motion practically coincides with corresponding derivative $\partial\epsilon/\partial k$, while direction of motion is defined by the sign of this derivative (notice that swapping the order of arrays across the interface would result in inversion of the direction of motion). Small-amplitude waves remaining behind quasi-soliton and visible in distributions at different moments of time are bound to the interface and appear because higher-order dispersion effects are unavoidable in this system and come into play for relatively well-localized states, like the one excited in Fig. 7(b). However, as one can see from amplitude dependencies in Fig. 6(c), such radiation does not lead to noticeable decrease of amplitude even at times $t\sim 10^3$, i.e. interface quasi-solitons are exceptionally robust objects. Quasi-soliton illustrated in Fig. 7(b) has most of its norm concentrated within honeycomb array. For other properly selected values of momentum one may also excite quasi-solitons with practically equal norm sharing between two arrays or solitons concentrated mainly within Lieb array.

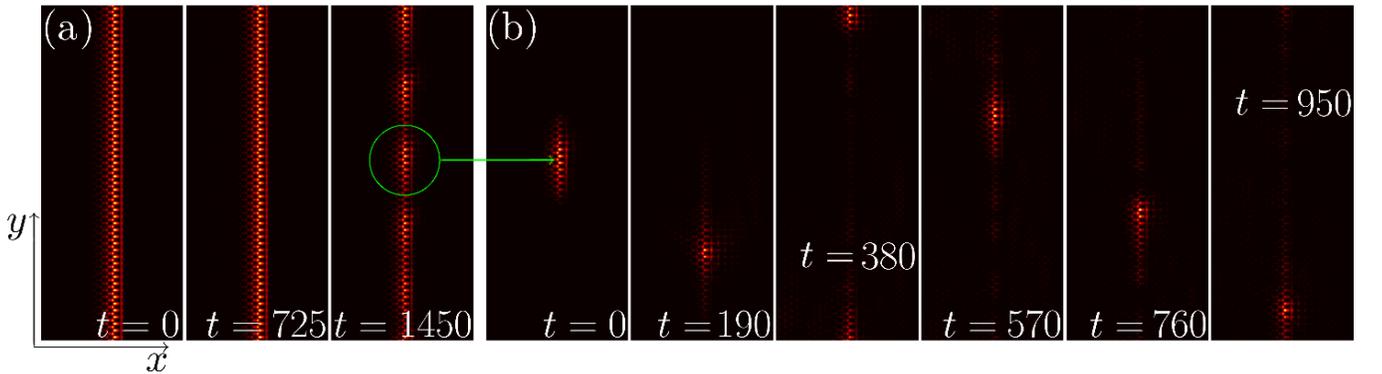

Fig. 7. (a) Distributions of $|\psi_-|$ in different moments of time corresponding to the green circles in Fig. 6(b) in the perturbed nonlinear interface state. (b) Unidirectional motion of quasi-soliton obtained from the bright spot marked with a green circle in (a). All distributions are shown within $-10\le x\le 40$ and $-50\,\text{Y}\le y\le +50\,\text{Y}$ windows. All quantities are plotted in dimensionless units.

## 6. The impact of losses and coherent pump

Polariton condensates are inherently dissipative, hence the impact of losses on edge states should be considered. Such losses can be compensated by the external pump that can be coherent or incoherent. In this section, we consider coherent pump. We show that nonlinear edge states can be excited by such a pump and now they stem from the balance of not only dispersion and nonlinearity, but also from balance between pump and losses. Importantly, coherent pump offers additional tools for control of the propagation direction and density distribution in edge states.

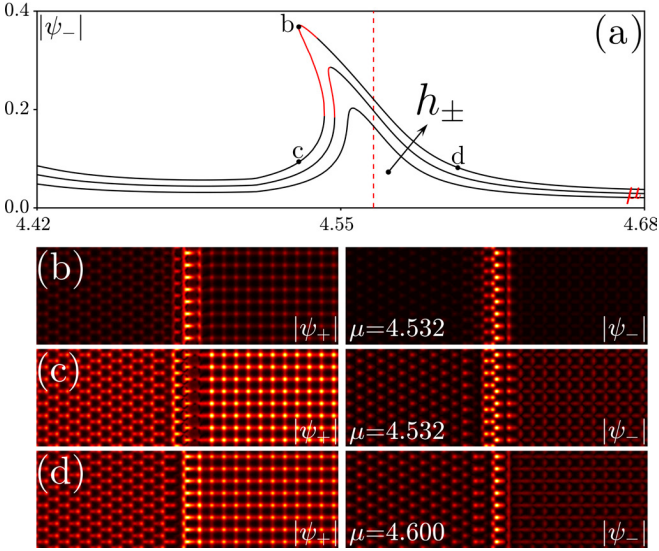

Fig. 8. (a) Peak amplitude of the dominating spin-negative edge state component versus $\mu$ for the pump amplitude $h_\pm = 0.0025$, $0.0035$, and $0.0045$ at $k = 0.3K$, $\gamma = 0.008$. Arrow indicates the direction of the increase of the pump amplitude. Stable and unstable states are shown black and red, respectively. Vertical red dashed line marks the energy of the linear pump-free edge state. (b)-(d) Amplitude distributions of the spin-positive and spin-negative components, corresponding to the dots in (a).

The modified version of Eq. (1) accounting for pump and losses can be written as [51]:

$$i\frac{\partial \psi_\pm}{\partial t} = -\frac{1}{2}\left(\frac{\partial^2}{\partial x^2} + \frac{\partial^2}{\partial y^2}\right)\psi_\pm + \beta\left(\frac{\partial}{\partial x} \mp i\frac{\partial}{\partial y}\right)^2 \psi_\mp \pm \Omega\psi_\pm + \mathcal{R}(x,y)\psi_\pm + (|\psi_\pm|^2 + \sigma|\psi_\mp|^2)\psi_\pm - i\gamma\psi_\pm + \mathcal{H}_\pm(y,t). \quad (5)$$

Here we assume equal losses $\gamma = 0.008$ in two spinor components and polarized pump $\mathcal{H}_\pm = h_\pm \exp(iky + i\mu t)$, where $h_\pm$ is the pump amplitude. We assume that the pump is periodic in the $y$-direction and that its frequency determines the value of detuning $\mu$. Below we consider a linearly polarized pump, $h_+ = h_-$, and fix the pump momentum to $k = 0.3K$ [there is only one interface state in the top gap for this momentum value; please see Fig. 1(f)]. Changing the detuning $\mu$ within the topological gap leads to resonant excitation of the interface state, when detuning $\mu$ matches the energy of the linear interface state in corresponding conservative system [see dashed line in Fig. 8(a)]. However, because our system is nonlinear, resonance tilts that may lead to bistability (coexistence of several states) for sufficiently large pump amplitudes. Typical resonance dependencies of the amplitude of the spin-negative component on detuning $\mu$ are shown in Fig. 8(a), clearly revealing bistability. It should be stressed that localization degree of the excited interface state dramatically depends on detuning $\mu$ and pump momentum $k$. In Figs. 8(b)-8(d), the distributions $|\psi_\pm|$ are displayed for several detuning values that are marked by dots in Fig. 8(a). In Figs. 8(b) and 8(c), the energies are same but the localization of the two interface states is considerably different. The state that is close to the resonance tip is strongly localized on the interface [Fig. 8(b)]. Localization rapidly decreases as one moves away from resonance, see Fig. 8(d). We also checked stability of the interface states in the presence of pump and losses. For large pump amplitudes the states from the upper branch of resonance curves close to the tip may be unstable, but largest part of the upper branch is stable.

## 7. Conclusions

Summarizing, we have predicted that the interface between honeycomb and Lieb arrays of polariton microcavity pillars can support linear and nonlinear interface states, whose number and localization properties in two different arrays can be controlled by various means, including modification of array depths, Bloch momentum, and nonlinear effects. We obtained unidirectional interface quasi-solitons as a result of development of modulation instability for nonlinear edge states. Bistability of the nonlinear interface states in the presence of coherent pump is demonstrated too. Our findings may be extended to various photonic and matter-wave settings involving interfaces between two different topological periodic media.

**Acknowledgements:** Y.Q.Z. acknowledges support by National Key R&D Program of China (2017YFA0303703), Natural Science Foundation of Guangdong province (2018A0303130057) and Shaanxi province (2017JZ019, 2016JM6029). Y.V.K. acknowledges funding of this study by RFBR and DFG according to the research project № 18-502-12080. A.F. gratefully acknowledges support by Spanish MINECO through project no. TEC2017-86102-C2-1-R. This work was partially supported by the program 1.4 of Presidium of RAS "Topical problems of low temperature physics."